\documentstyle[12pt]{article}
\date{}
\begin{document}
\rightline{CU-TP-941}
\vskip 20pt
\begin{center}
{\Large\bf Toward Equilibration in the Early Stages\\
 After a High Energy
Heavy Ion Collision}
\vskip20pt

{A.H.Mueller\footnote{This work is supported in part by the US
Department of Energy, grant DE-FG02-94ER-40819}\\
Physics Department, Columbia University\\
New York, New York 10027}.
\end{center}

\vskip 20pt
\noindent{\bf Abstract}
\vskip 15pt
The early stages in the evolution of the gluons produced in the central region of a head-on high-energy heavy ion
collision are studied. An equation is given for the rate of change of transverse momentum into longitudinal momentum where 
the longitudinal direction is along the collision axis. We are able to follow the system up to the time where equilibration seems to be setting in, but we
are unable to actually follow the system as it reaches equilibrium.

\section{Introduction}

At very small values of \ x\ the gluon density in a proton's light-cone wavefunction reaches
saturation\cite{Gri,Mue,Jal,ler}, that is for gluons having transverse momentum below the saturation momentum,
$Q_s,$ there is a density on the order of ${(N_c^2-1)\over \alpha N_c}$ gluons per unit of transverse phase space
(transverse coordinate times transverse momentum space).  Parton saturation in the proton may already have been
observed at HERA\cite{Cal,Abr}.  This regime of high density gluons is reached at more moderate values of \ x\
for a large nucleus.  In a head-on collision of high energy heavy ions saturated gluons, those having transverse
momentum at or below $Q_s,$ are freed at very early times after the valence quarks of the two ions have passed each
other, say in the center of mass frame.  It is widely assumed that this gluon system then quickly reaches kinetic
equilibrium and somewhat more slowly reaches equilibrium between gluons and quarks.

There are Monte Carlo calculations which have followed the early stages of the evolution of a heavy ion collision
and which give good evidence for equilibration\cite{Kin,Zha}.  There is also an interesting Monte Carlo
calculation\cite{Kra} which followed the time evolution of the gluon fields which start from the saturated
distribution occurring in the McLerran-Venugopalan model.  Our purpose  here is somewhat different.  The object is
to give a rough analytic calculation of how gluons evolve in momentum and in space after being freed in a high
energy heavy ion collision. We are able to follow a typical gluon through its many small angle scatterings up to
the point where the interactions which should actually give equilibration begin to occur.  At the time where
equilibration actually begins to set in our technique breaks down, although we are able to say at what time and
at what momentum, relative to the saturation momentum, equilibration starts to set in.

The starting point of our discussion is the McLerran-Venugopalan\cite{McL} model for the light-cone wavefunction
of a heavy ion\cite{Kov}.  This model is characterized by one parameter the saturation momentum, $Q_s,$ below
which gluon densities reach their maximum value.  We suppose that a finite fraction (our parameter\  $c$\  in
(16)) of all gluons having transverse momentum less than or equal to $Q_s$ are freed in a head-on heavy ion
collision\cite{Bla}, and that those gluons having small rapidity are freed in a time about equal to $1/Q_s.$ 
Gluons having small rapidity can then scatter only with other gluons having small rapidity,  because the rate
of scattering is slow and higher rapidity gluons quickly separate from lower rapidity gluons.  It is then
straightforward to write an equation for the rate of change of the transverse momentum as zero rapidity gluons
scatter.  The equation is given  in (49).  In arriving at (49) we have also had to determine where the infrared
cutoff lies for small angle scattering.  We have done this by requiring that the soft gluon field, coming from the
hard gluons we are dealing with, also be limited in magnitude to be no larger than $1/g.$  This determines a
minimum m
\\omentum transfer,
$\ell_m,$ which is not quite the same as has been suggested earlier\cite{Bir}.  The difference between $\ell_m$
given in (44) and the screening length of Ref.13, given in (45), changes the constants in (49), but not the form
of the equation.

What has been said for zero rapidity gluons also applies to gluons having rapidity $y,$ by simply considering
their interactions in a frame boosted by -$y.$  Thus gluons of rapidity\  $y$\  only interaction with other gluons
having rapidity $y$ and (49) remains valid for the rate of change of transverse momentum if $\xi \to \xi +
\ell\ n\ ch\ y.$

At early times the rate of change of transverse momentum is much less than that corresponding to the
one-dimensional expansion of a gas of gluons in equilibrium.  However, because of the time dependence of the
infrared cutoff the rate of loss of the square of the transverse momentum of gluons increases so that at a time,
for zero rapidity gluons, on the order of ${1\over Q_s}\  e^{{\sqrt{{2\pi \over c\alpha N_c}}}},$ and at a
transverse momentum given by $(\ell_\perp/Q_s)^2 \sim {\sqrt{{c\alpha N_c\over 2\pi}}},$ the expansion of our
gluon gas more closely resembles that of a system in equilibrium.  At this same time scatterings which change
longitudinal momentum into transverse momentum, not included in (49), begin to become important. This is also the
time at which gluon interactions no longer take place simply among gluons of the same rapidity so that our whole
formalism is breaking down.  Finally, this is the time at which the $z-$extent of the medium is becoming greater
than the mean free path for gluon interactions which can change a gluon's momentum by an amount comparable to its
momentum.  It seems very likely that this is the time at which equilibration is setting in, although we are
unable to follow the system further because of the limitations of our present formalism. It is important to note
that the time when equilibration seems to be setting in is parametrically short compared with the radius of the
initial colliding ions. We consider this evidence that equilibrium indeed occurs in very high energy relativistic
heavy ion collisions while the system is still undergoing a one-dimensional expansion.

\section{Small-x wavefunctions and parton saturation}

In this section we shall review the parton description of the light-cone wavefunction of a heavy ion.  There is
a very simple model, the McLerran-Venugopalan\cite{McL} model, which very nicely illustrates parton (gluon)
saturation.  Indeed, it has recently been claimed\cite{ler}  that the main results of the McLerran-Venugopalan
model should be general results in QCD.  We shall briefly review the physical basis of this model and then
comment on the general expectation of the reliability of the model.

We begin, following McLerran and Venugopalan, by looking at the distribution of valence quarks in a high energy
ion.  The valence quarks are found in a Lorentz-contracted longitudinal disc of size $\Delta\ z =2R\cdot{m\over
p}$ where  $R$\  is the nucleon size and \ $p$\ the momentum per nucleon of the ion.  For purposes of our
qualitative discussion we consider $\Delta \ z = 0$ so that one can imagine the valence quarks having a
two-dimensional number density of

\begin{equation}
n_q(b) = 6\rho{\sqrt{R^2-b^2}}
\end{equation}

\noindent valence quarks per unit area where\  $\rho$\  is the normal nuclear density while \ $b$\ is the impact
parameter measured from the center of the nucleus in a direction perpendicular to the direction of motion of the
nucleus.

The valence quark density given in (1) is the source of soft gluons corresponding to the Weizs\"acker-Williams
field of $n_q$\cite{Jal,Yu}.  The color charge at a given impact parameter comes from a random addition of the
color charges of each of the valence quarks at that impact parameter. A single quark gives ${\alpha C_F\over \pi}\
\ell  n Q^2/\mu^2$ gluons at scale $Q^2$ and per unit rapidity so that one expects the number density of gluons per
unit area and per unit rapidity to be

\begin{equation}
{d x G_A(x,Q^2)\over d^2b}\ =\ n_q(b) {\alpha C_F\over \pi}\ \ell n\ Q^2/\mu^2
\end{equation}

\noindent leading to a gluon distribution in the nucleus of
\begin{equation}
x G_A(x,Q^2) = 3A {\alpha C_F\over \pi} \ell n\ Q^2/\mu^2\ = A xG(x,Q^2).
\end{equation}

\noindent (We note that for a proton (3) is not too bad when $10^{-1} < x < 10^{-2}$ and $Q^2 = 5-10 GeV^2.)$  The
unitegrated gluon distribution

\begin{equation}
{dxG_A(x,\ell^2)\over d^2bd^2\ell_\perp}\ =\ n_q(b) {\alpha C_F\over \pi^2\ell_\perp^2}
\end{equation}

\noindent has the interpretation, in light-cone gauge, of the number of gluons per unit rapidity per unit of
transverse phase space (impact parameter space times transverse momentum space).  Of course (2)-(4) are correct
only in the low-density limit.  In the high density regime the distribution of gluons in the wavefunction of the
nucleus in the McLerran-Venugopalan model is\cite{Cal,Kov}

\begin{equation}
{dxG\over d^2bd^2\ell_\perp} = {N_c^2-1\over 4\pi^4\alpha N_c}\ \int {d^2x_1\over  
\underline{x}^2}(1-e^{-\underline{x}^2Q_s^2/4}) e^{i\underline{\ell}\cdot \underline{x}}
\end{equation}

\noindent where the saturation momentum, $Q_s,$ is given by

\begin{equation}
Q_s^2 = {8\pi^2\alpha N_c\over N_c^2-1} {\sqrt{R^2-b^2}}\ \rho x G(x,{4\over \underline{x}^2}).
\end{equation}

\noindent The large $\ell^2$ limit given in (4) comes from the logarithmic singularity of $xG$ when evaluating
(5).  In the region $\ell_\perp^2/Q_s^2 <<1$  one can neglect the $\underline{x}^2$ dependence of $Q_s^2$ and
write

\begin{equation}
{dxG\over d^2bd^2\ell_\perp} = {N_c^2-1\over 4\pi^3\alpha N_c}\ \int_1^\infty {dt\over t}\ e^{-t\ell_\perp^2/Q_s^2}
\end{equation}

\noindent with the scale of $xG$ in (6) taken as ${4\over \underline{x}^2} = Q_s^2.$  From (7) one finds, when 
$\ell_\perp^2/Q_s^2 <<1,$

\begin{equation}
{dxG\over d^2bd^2\ell_\perp} = {N_c^2-1\over 4\pi^3\alpha N_c}\ \ell\ n\ Q_s^2/\ell_\perp^2
\end{equation}

\noindent which exhibits saturation, the logarithm naturally present in the momentum space expression.  In Ref.4 a
result essentially identical to (8) has been derived using BFKL dynamics.  We expect the essential features of
(7), and (8), to be general though it is difficult to reliably set the overall constant in (8).

Despite our notation it should be emphasized that (8) refers to the distribution of gluons in a light-cone
wavefunction.  It has no relation to the gluon distribution, or its derivatives, as determined by an operator
product expansion analysis of deep inelastic lepton-nucleus scattering.  The result in (8) does, however, give
the transverse momentum distribution of gluons produced in a hard scattering off a large nucleus.

Of course, the whole approach discussed here only makes sense if the parameter $Q_s^2$ is in the perturbative
regime.  For RHIC energies (6) gives $Q_s\approx 1 GeV$ while for LHC energies $Q_s$ should be $2-3 GeV.$  Thus,
our perturbative approach is certainly marginal for central RHIC collisions while it should be more applicable for
heavy ion collisions at the LHC.  Of course the same phenomenon of saturation also occurs at very small $ x$ 
values for a proton and, indeed,  gluon saturation may have been observed at HERA.

\section{Toward equilibration in the central region of a head-on heavy ion collision.}

In this section we shall discuss, in a semiquantitative manner, the evolution from very early times up to the
time where equilibrating interactions begin to occur for the central region of a zero impact parmeter heavy ion
collision.  There is good evidence from Monte Carlo simulations that equilibration does indeed occur in collisions
having kinematics similar to those which will take place at RHIC and at the LHC\cite{Kin,Zha}.Our purpose here is
of a more theoretical nature, to try and understand in analytic detail the early stages of evolution of a dense
gluon system and to see that equilibrating interactions do begin well before the system falls apart.

\subsection{Which gluons are freed and when?}

At the time of collision of the two ions the gluons in the wavefunctions of the ions are distributed according to
(5).  For $\ell^2/Q_s^2 <<1$ the distribution is given by (8), a distribution which is, except for a logarithm,
given by phase space.  Since there are few gluons in the wavefunctions having $\ell_\perp^2/Q_s^2<<1$ we shall
simply ignore them completely in determining the initial distribution of gluons.  When $\ell_\perp^2/Q_s^2 >>1$
the distribution of gluons in the wavefunctions is given by (2) and (4).  These gluons are additive in the various
nucleons in the nucleus so that one may write the two gluon (jet) production cross section as

\begin{equation}
{d\sigma\over dy_1dy_2d^2\ell_\perp\ d^2b} = \int\  d^2b_1d^2b_2 {dx_1 G_A(x_1\ell^2)\over d^2b_1}\ 
{dx_2G_A(x_2,\ell^2)\over d^2b_2}\  {d\hat{\sigma}\over d\hat{t}}\ 
\delta(\underline{b}_1-\underline{b}_2-\underline{b})
\end{equation}

\noindent where $xG_A$ is given by (2) and where the gluon-gluon elastic scattering cross section is

\begin{equation}
{d\hat{\sigma}\over d\hat{t}} = ({\alpha N_c\over \pi})^2\  {4\pi^3\over (N_c^2-1)\hat{s}^2} (3-{\hat{u}\hat{t}\over
\hat{s}^2}\  - {\hat{u}\hat{s}\over \hat{t}^2} \ - {\hat{s}\hat{t}\over \hat{u}^2})
\end{equation}

\noindent with $\hat{t}= - \ell_\perp^2.$  The main interest of (9) and (10) for us here is that for
$\ell_\perp^2/Q_s^2 >> 1$ the cross section in (9) scales like $\ell_\perp^{-4}.$   Thus one gets more particles,
and more transverse energy, by lowering $\ell_\perp.$  Thus the dominant region for gluon production, and for the
production of transverse energy, in a very high energy heavy ion collision is the region where the produced gluons
have transverse momentum on the order of $Q_s,$\cite{Bla} and where the gluons come from the interaction of gluons
having transverse momenta on the order of $Q_s$ in the wavefunctions of the colliding ions.

Now that it is clear that the main contribution to the initial transverse energy density in a high energy heavy
ion collision comes from gluons having transverse momentum on the order of $Q_s,$ the saturation momentum, we need
to ask at what time are these gluons produced.  Suppose we focus on a unit of rapidity centered about $Y_0, -
{1\over 2} + Y_0 < y < {1\over 2} + Y_0$ with quanta in this region being right-movers in the frame of choice. 
Then define $\hat{P}$ by

\begin{equation}
\hat{P} = Q_s \ sinh\  Y_0.
\end{equation}

\noindent $\hat{P}$ is the typical z-component of the momentum of the gluons we are considering.  The time over
which the gluons in our unit of rapidity are freed is

\begin{equation}
\hat{\tau} = {2\hat{P}\over Q_s^2}.
\end{equation}

\noindent $\hat{\tau}$ is the time over which gluons in differing units of rapidity physically separate with those
we are considering.  It is also the time over which our gluons physically separate from the valence quarks.  We may
estimate the time of this latter separation by neglecting the z-extent of the right-moving valence quarks.  The
z-extent of the gluons in $-{1\over 2} + Y_0 < < y < {1\over 2} + Y_0$ is

\begin{equation}
\Delta z = {1\over \hat{P}}
\end{equation}

\noindent while their longitudinal velocity is

\begin{equation}
v = {\hat{P}\over {\sqrt{\hat{P}^2+Q_s^2}}}\approx 1 - {Q_s^2\over 2\hat{P}^2}.
\end{equation}

\noindent The time which it takes the right-moving valence quarks to separate from the gluons in our unit of
rapidity is

\begin{equation}
\Delta \tau = {\Delta z\over 1-v}.
\end{equation}

\noindent Using (13) and (14) we see that $\Delta \tau = \hat{\tau}.$

Thus the gluons on which we are focusing are freed during a time given by (12).  We have described these gluons as
being virtual gluons in the light-cone wavefunction which are "freed" during the collision, and this is indeed the
picture we have in mind.  However, we do not preclude that a significant number of these gluons may be created, in
a light-cone gauge description, after the valence quarks of the colliding nuclei passed each other. In this case it
is more proper to describe these gluons as \underline{produced} by the collision.

It would be nice to have a reliable calculation of the production cross section for freeing gluons in a central
ion-ion collision.  Unfortunately, this has not been achieved, up to now, even in the McLerran-Venugopalan model. 
There is an interesting numerical calculation of some of the features of early time production\cite{Kra}, but so
far it is not clear how to convert these calculations into an evaluation of the number of gluons produced at time
$\hat{\tau}$  In order to proceed further we simply suppose that all the gluons in the light-cone wavefunction
having $\ell^2 \approx Q_s^2$ are freed.  Thus we take as our initial distribution of freed gluons

\begin{equation}
{dN\over d^2bdy} = c^\prime{dxG\over d^2bd^2\ell}\bigg{\vert}_{\ell^2=Q_s^2}\ \pi Q_s^2 \ = c {N_c^2-1\over
4\pi^2\alpha N_c}\ Q_s^2
\end{equation}

\noindent where \ $c$\ is a constant expected to be of order one.  \ $c$\ cannot be calculated without knowing
\underline{exactly} which gluons are freed during the collision.  At early times we can use free streaming
kinematics

\begin{equation}
z = t \ tanh\ y
\end{equation}

\noindent to get

\begin{equation}
{dN\over d^2bdz} = c {(N_c^2-1)Q_s^2\over 4\pi^2\alpha N_c} \ {t\over t^2-z^2}.
\end{equation}

\noindent Later on we shall argue that (18) should be a reasonable approximation for all times well before
equilibration takes place.

\subsection{The dominant two-body scatterings at early times}

Immediately after the gluons in a given rapidity region are freed they begin to move freely along straight line
trajectories until scattering with other freed gluons makes them deviate from their original direction. We suppose
that the main scattering which occurs is gluon-gluon elastic scattering.  Such an assumption should be justified
for sufficiently large $Q_s.$  Consider those gluons having zero rapidity in a given frame of reference, where we
suppose the z-direction is the colinear direction of the colliding ions. If gluons $\ell_1$ and $\ell_2$ scatter
into gluons $\ell_1^\prime$ and $\ell_2^\prime$ the kinematics of the scattering can be represented as

\begin{equation}
\ell_{1\mu} = (\ell_{10},\ell_{1x},\ell_{1y},\ell_{1z}) = \ell_1(1,1,0,0)
\end{equation}

\begin{equation}
\ell_{2\mu}=\ell_2(1,-cos \chi, -sin \chi, 0)
\end{equation}

\begin{equation}
\ell_{1\mu}^\prime = \ell_1^\prime (1, cos \theta, sin \theta cos \varphi, sin \theta sin \varphi)
\end{equation}

\begin{equation}
\ell_{2\mu}^\prime = (\ell_1 + \ell_2-\ell_1^\prime)_\mu
\end{equation}

\noindent with $\ell_1^\prime$ determined by $(\ell_2^\prime)^2 = 0.$  The initial gluons lie in the $x,y$ plane
and approach each other at an angle $\chi$ with the initial gluon $\ell_1$ being scattered by an angle $\theta$
from its original direction into the final gluon $\ell_1^\prime.\ \   \varphi$ is an azimuthal angle, but here about
the x-axis.  Useful invariants are

\begin{equation}
\hat{s} = (\ell_1+\ell_2)^2 = 2\ell_1\ell_2(1\ +\ cos\  \chi)
\end{equation}

\begin{equation}
-\hat{t}  = - (\ell_1-\ell_1^\prime)^2 = 2\ell_1 \ell_1^\prime(\ 1\ -\ cos\  \theta)
\end{equation}

\noindent while $(\ell_1+\ell_2-\ell_1^\prime)^2 = 0$ gives

\begin{equation}
\ell_1^\prime = {\ell_1\ell_2(1\ +\ cos\  \chi)\over \ell_1(1\ -\ cos\  \theta) + \ell_2(1+ cos \chi cos \theta\  +
sin\ 
\chi\  sin\ 
\theta\  cos\  \varphi)}.
\end{equation}

\noindent We shall be dealing with small angle scattering ($\theta$ small) so that

\begin{equation}
\ell_1^\prime = {\ell_1\ell_2(1\ +\  cos\  \chi)\over \ell_2(1\  +\  cos\  \chi) + \ell_2 \theta\  sin\  \chi\ 
cos\ 
\varphi + {\theta\over 2}^2(\ell_1-\ell_2 cos\  \chi)}.
\end{equation}

In what follows we shall be following the change in transverse momentum of a gluon due to scattering with other
gluons in the medium. For the gluon initially labeled by $\ell_1$ the transverse momentum is $ \ell_{1\perp}=
{\sqrt{\ell_{1x}^2 + \ell_{1y}^2}} = \ell_1.$  After scattering

\begin{equation}
\ell_{1\perp}^\prime = {\sqrt{(\ell_{1x}^\prime)^2+ (\ell_{1y}^\prime)^2}} = \ell_1^\prime(1-{\theta^2\over 2}
sin^2\phi)
\end{equation} 

\noindent in the small $\theta$ approximation.  Thus

\begin{equation}
\ell_{1\perp}-\ell_{1\perp}^\prime = \ell_1-\ell_1^\prime + \ell_1 {\theta^2\over 2}\ sin^2\phi
\end{equation}

\noindent in the small $\theta$ approximation.  Using (26) it is straightforward to arrive at

\begin{eqnarray}
\ell_{1\perp}-\ell_{1\perp}^\prime &=& {\ell_1\theta^2\over 2\ell_2(1 + cos \chi)^2} [(\ell_1-\ell_2 cos \chi)(1 +
cos \chi) \nonumber \\
&-& 2\ell_2 sin^2 \chi   cos^2\varphi  + \ell_2(1 + cos \chi)^2 sin^2\phi].
\end{eqnarray}

\noindent After averaging over the azimuthal angle $\phi$ according to

\begin{equation}
\overline{\Delta\  \ell_{1\perp}}\equiv \int_0^{2\pi} {d\phi\over 2\pi} \Delta \ell_{1\perp} \equiv \int_0^{2\pi}
{d\phi\over 2\pi} (\ell_{1\perp}-\ell_{1\perp}^\prime)
\end{equation}

\noindent one finds

\begin{equation}
\overline{\Delta \ell_{1\perp}} = {\ell_1\theta^2\over 4\ell_2(1+ cos \chi)} [2\ell_1-\ell_2(1-cos \chi)].
\end{equation}

\subsection{The rate of change of gluon transverse momentum at early times}

Now that we know how much transverse momentum a gluon loses during an early-time collision, given by (29) and
(31), it is not difficult to give an expression for the rate of change of the transverse momentum of a zero
rapidity gluon due to its initial small angle scatterings.  The result is

\begin{equation}
{d\ell_{1\perp}\over dt} = - \int {dN\over d^3x d^2\ell_2} \vert \vec{v}_{12}\vert {d\sigma\over
d\Omega_1^\prime}\cdot \Delta \ell_{1\perp} d\Omega_1^\prime d^2\ell_2
\end{equation}

\noindent where

\begin{equation}
d\Omega_1^\prime = sin \theta d\theta  d \varphi
\end{equation}

\begin{equation}
\vert \vec{v}_{12}\vert = {\sqrt{(\vec{v}_1-\vec{v}_2)^2 - (\vec{v}_1x\vec{v}_2)^2}}.
\end{equation}

\noindent Eq.32 expresses the fact that the rate of change of transverse momentum is given by the change of
transverse momentum in a single scattering event times the rate of scatterings.  The rate of scatterings is given
by the cross section times the  relative flux, $\vert \vec{v}_{12}\vert,$ times the density of scatterers.  The
density of scatterers is given by (18) as

\begin{equation}
{dN\over d^3xd^2\ell_2} = {\Theta(Q_s^2-\ell_2^2)\over \pi Q_s^2}\ {dN\over d^2bdz} = c {(N_c^2-1)\over
4\pi^3\alpha N_ct} \Theta (Q_s^2-\ell_2^2)
\end{equation}

\noindent where we approximate the $\ell_2-$spectrum by a step function.  $\vert \vec{v}_{12}\vert$ can be
evaluted from (19), (20) and (34) as

\begin{equation}
\vert \vec{v}_{12}\vert = 1 + cos \chi
\end{equation}

\noindent while, for small angle scattering, using (10) and (24),

\begin{equation}
{d\sigma\over d\Omega_1^\prime} d\Omega_1^\prime = ({\alpha N_c\over \pi})^2 {4\pi^3\over N_c^2-1}\
{d\theta^2\over \ell_1^2\theta^4}\ {d\varphi\over 2\pi}.
\end{equation}

\noindent The $d\varphi$ in (37) and the lack of any $\varphi$ dependence in any of the other factors, beside
$\Delta \ell_{1\perp},$ in (32) allow one to replace $\Delta \ell_{1\perp} {d\varphi\over 2\pi}$  by
$\overline{\Delta} \ell_{1\perp}$ as given by (31).  Thus

\begin{equation}
{d\ell_{1\perp}^2\over dt}\ =\ - c\ {\alpha N_c\over \pi t}\ \int {d\theta^2\over \theta^2} \int^{Q_s}
d\ell_2 (2\ell_1-\ell_2)
\end{equation}

\noindent or

\begin{equation}
{d\ell_{1\perp}^2\over d t} =\ -\ {3 c \alpha  N_c\over 2\pi t} \ Q_s^2 \int_{\theta_m^2}^1 {d\theta^2\over \theta^2}\
=\ - {c\alpha N_c\over \pi t} Q_s^2\   \int_{\theta_m^3}^1\  {d\theta^3\over \theta^3},
\end{equation}

\noindent where  the $\theta-$integration is cut off at $\theta_m,$ which quantity will be the topic of our next
section.  Finally, supressing the subscript  1  on $\ell_{1\perp},$  one finds

\begin{equation}
{d\ell_\perp^2\over dt}\ =\ -\ {3 c \alpha N_c\over 2\pi t} \ Q_s^2\ \ell n\  1/\theta_m^2,
\end{equation}

\noindent a formula accurate in the logarithmic approximation.

\subsection{The minimum scattering angle}

The cutoff on the $\theta-$ integration in (39) corresponds to a cutoff on the momentum transfer squared,
$\hat{t},$ given by

\begin{equation}
- \hat{t}_m\ = \ \ell_{1\perp}^2 \theta_m^2  =  \ell_m^2
\end{equation}

\noindent with $\ell_m$ the minimum momentum transfer allowed.  One can view (39) as arising from the scattering
of a particular hard gluon with the (relatively) soft gluon field due to all the other hard gluons which have been
produced at a similar rapidity.  However, this soft field should also be limited to size  $1/g$ because of
saturation.  We can impose the saturation condition by requiring that our hard gluon have only one scattering
with momentum transfer greater than or equal to $\ell_m$ in a distance equal to $1/\ell_m.$  Thus, $\ell_m,$ and
hence $\theta_m$ is determined by

\begin{equation}
\int_{-\infty}^{-\hat{t}_m}{d\sigma\over d\hat{t}}\ \cdot\ {dN\over d^3x}\ \cdot\ {1\over \ell_m}\ =\ 1.
\end{equation}

\noindent Using (10), (18) and (41) one finds

\begin{equation}
\theta_m^3 = c({\alpha N_c\over \pi})\ {Q_s^3\over \ell^3_{1\perp}(Q_st)}
\end{equation}

\noindent or

\begin{equation}
\ell_m^3 = c({\alpha N_c\over \pi})\ {Q_s^3\over Q_st}.
\end{equation}

\noindent This is not quite the same screening mass suggested in Ref.13.  If one uses Eq.3 of Ref.13 with
${dN_{AA}\over d^2\ell dy}$ given from (16) above and with all freed gluons taken in the integral of Eq.3 one
finds

\begin{equation}
\mu_D^2 = 2c {Q_s^2\over Q_st}.
\end{equation}

\noindent When $(\alpha N_c)^2 Q_st$ is much bigger than one $\ell_m,$ as given by (44), is larger than $\mu_D,$
as given in (45).  For our purposes the exact form of the cutoff in the infrared is not so important so long as
the cutoff decreases no faster than some fractional power of $Q_st.$

\subsection{The time dependence of $\ell_\perp$}

Now go back to (40).  It is convenient to write this equation as 

\begin{equation}
{1\over Q_s^2} \ {d\ell_\perp^2(t)\over dt}\ =\ - {c \alpha N_c Q_s\over \pi Q_st}\ \ell n\ {\ell_\perp^3(t)\over
\ell_m^3}.
\end{equation} 

\noindent Using (44),

\begin{equation}
{d(\ell_\perp^2(t)/Q_s^2)\over d\ell n Q_st}\ =\ - c {\alpha N_c\over \pi} \left[{3\over 2} \ell n\
{\ell_\perp^2\over Q_s^2} + \ell n\ Q_st\ +\ \ell n({\pi\over c \alpha N_c})\right].
\end{equation}
\noindent Defining

\begin{equation}
\xi = \ell n\ Q_st,\  \xi_0 =\ \ell n ({c \alpha N_c\over \pi})
\end{equation}
\noindent one finds

\begin{equation}
{d(\ell_\perp/Q_s)^2\over d\xi}\ =\ - c{\alpha N_c\over \pi} \left[ \xi - \xi_0 + 3 \ell n(\ell_\perp/Q_s)\right]
\end{equation}

\noindent which is our final equation.

So far we have derived (49) when $t,$ and hence $\xi,$ is not too large, and thus in the regime where
$\ell_\perp/Q_s$ is very close to 1.  In the next section we shall argue that (49) is valid in a much wider range
of times.  Thus, we here give a brief discussion of the properties of the solution to (49) even when $\xi$
corresponds to very large times.  We are interested in finding the solution to (49) in the region $\xi > 0$ with
$q_\perp/Q_s = 1$ at $\xi = 0.$  The dependence of $\ell_\perp/Q_s$ on $\xi$ is qualitatively clear.  For $\xi$
not too large one may neglect the $\ell n \ell_\perp/Q_s$ term on the right-hand side of (49) to get

\begin{equation}
(\ell_\perp/Q_s)^2 \approx 1 - {c\alpha N_c\over 2\pi} \left[(\xi - \xi_0)^2-\xi_0^2\right].
\end{equation}

\noindent $\ell_\perp/Q_s$ continues to decrease with increasing $\xi.$  When $\xi$ is large, and hence
$\ell_\perp/Q_s$ is small, the $\xi-$dependence of $\ell_\perp/Q_s$ is obtained by setting the right-hand side
of (49) equal to zero.  This gives

\begin{equation}
(\ell_\perp/Q_s)^2 _{\longrightarrow\atop{\xi-\xi_o\to \infty}}e^{-{2\over 3}(\xi-\xi_0)}.
\end{equation}

\noindent The transition from (50) to (51) takes place when $\xi$ is extremely close to $\xi_1$ with $\xi_1$ given
by

\begin{equation}
\xi_1 = {\sqrt{{2\pi\over c \alpha N_c}+ \xi_0^2}} + \xi_0
\end{equation}
\noindent or, equivalently, when $t$ is near $t_1$ given by

\begin{equation}
t_1 = c {\alpha N_c\over \pi Q_s} e^{{\sqrt{{2\pi\over c \alpha N_c}+ \xi_0^2}}}.
\end{equation}

Equation (49) represents scatterings which transform transverse momenta into longitudinal momenta. At early times
there is a negligible transfer of longitudinal momenta into transverse momenta.  As we shall see in the next
section it is exactly at the time when longitudinally moving particles have a significant probability of
scattering and creating additional transverse momenta that equilibration is beginning to set in.

There is another way to estimate when equilibration begins to set in.  An equilibrated expanding gas in QCD should
obey the ideal gas expansion law

\begin{equation}
\left({T\over T_0}\right)\ =\ \left({t\over t_0}\right)^{-1/3}
\end{equation}

\noindent with  $T$  the (large) temperature of the gas.  For an equilibrated system transverse momentum and
temperature are proportional so one should have

\begin{equation}
{-d(\ell_\perp/Q_s)^2\over d\xi}\ =\ {2\over 3} (\ell_\perp/Q_s)^2.
\end{equation}

\noindent At early times, $\xi - \xi_0$ small, it is clear from (49) that the
transverse momentum in our expanding gluon system is not decreasing anywhere near fast enough to be close to
equilibration.  However, because of the lowering of $\ell_m$ as $t$ increases the rate of decease of
$\ell_\perp/Q_s$ grows with time and we may estimate the time when equilibration is likely to begin to happen by equating
(49) and (55).  (In the next section we shall see that this estimate is the same as that coming from requiring
that the mean free path of longitudinally moving gluons be less than the longitudinal length of the medium.)  One
finds

\begin{equation}
-{2\over 3} (\ell_\perp/Q_s)^2 =\ -\ c {\alpha N_c\over \pi}\left[\xi - \xi_0 + 3 \ell n\ \ell_\perp/Q_s\right].
\end{equation}

\noindent Taking $(\ell_\perp/Q_s)^2$ on the left-hand side of (56) from (50) and (temporarily) neglecting $\ell n\
\ell_\perp/Q_s$ one gets

\begin{equation}
1-c{\alpha N_c\over 2\pi}(\xi - \xi_0)^2 = {3 c \alpha N_c\over 2\pi} (\xi - \xi_0)
\end{equation}

\noindent giving $\xi = \xi_2$ with $\xi_2$ determined by

\begin{equation}
\xi_2 = {\sqrt{{2\pi\over c \alpha N_c}}} + \xi_0 - 3/2 = \xi_1 - 3/2
\end{equation}

\noindent and

\begin{equation}
\left(\ell_\perp/Q_s\right)_{\xi_2}^2\ =\ 3{\sqrt{{c \alpha N_c\over 2\pi}}}.
\end{equation}

\noindent One can now go back to (56) and check that neglect of the $\ell n\ \ell_\perp/Q_s$ term on the
right-hand side of that equation amounts only to a very small correction to $\xi_2.$

Thus, our {\em guess}  is that equilibration starts to set in at a time $\xi_2$ given by (48) and (58) and at that time
$(\ell_\perp/Q_s)^2$ is given by (59).  Also at $\xi_2,$ from (44),

\begin{equation}
\ell n (Q_s^2/\wedge^2) - \ell n (\ell_m^2/\wedge^2) = {2\over 3} (\xi - \xi_0).
\end{equation}

\noindent Using (58) one sees that at $\xi = \xi_2$

\begin{equation}
{1\over \alpha(Q_s^2)} = {1\over \alpha(\ell_m^2)} = O({1\over {\sqrt{\alpha}}})
\end{equation}

\noindent so that running coupling effects should not be important in the analysis we have given.

Finally, we note that

\begin{equation}
t_2 = {1\over Q_s} e^{\xi_2} = {c\alpha N_c\over \pi Q_s} exp\left[{\sqrt{{2\pi\over c \alpha N_c}}}-{3\over
2}\right].
\end{equation}

\noindent Using the fact that parametrically $Q_s^2 \propto {\alpha N_c\over N_c^2-1}  R$ with\ $R$\ the nuclear radius, 
equilibration can take place before the one-dimensional expansion changes to a three-dimensional expansion.

\subsection{From early times to equalibration}

Suppose we take as our initial reference frame the center of mass of the central ion-ion collision with the
right-moving valence quarks crossing the left-moving valence quarks at $t=0.$  At $\xi = 0$ gluons at zero
rapidity are freed and begin to interact.  At early times a gluon at zero rapidity can only interact with other
gluons having small rapidity because all other gluons separate quickly in the z-direction from zero rapidity
gluons.  A gluon having rapidity \ $y$\ has a longitudinal velocity $v_z=tanh y$ and can interact only with other
gluons having rapidity close to $y$\  because it, too, separates longitudinally from all other gluons.  Of
course if two gluons having rapidity \ $y$\ interact we can change to a coordinate system, boosted by rapidity
$-y$\ from our original system, where these gluons have zero rapidity and our calculation given above applies. 
Thus, our calculation for changes in transverse momentum, which is boost invariant, is also true for gluons
having rapidity \ $y.$  In particular, (49) remains valid for the rate of change of  transverse momentum of a
gluon having \ $y$\ if one simply identifies $\xi$ as $\ell n Q_st-\ell n ch y$ and $\xi_0$ as $\ell n ({c\alpha
N_c\over \pi}) - \ell n ch y$ with (49) being valid for $\xi > 0.$

To see when the approximation of scattering only between gluons having identical rapidities breaks down it is
convenient to introduce $\zeta$ by

\begin{equation}
\xi - \xi_0\ =\  {\sqrt{{2\pi\over c \alpha N_c} + \xi_0^2}}-\zeta.
\end{equation}

\noindent Then (49) and (50) become

\begin{equation}
{d(\ell_\perp/Q_s)^2\over d\zeta} = {c \alpha N_c\over \pi} \left[{\sqrt{{2\pi\over c \alpha N_c}}}-\zeta + 3 \ell
n(\ell_\perp/Q_s)\right]
\end{equation}

\noindent and

\begin{equation}
(\ell_\perp/Q_s)^2 = 2 {\sqrt{{c \alpha N_c\over 2\pi}}} \zeta - {c \alpha N_c\over 2\pi}  \zeta^2
\end{equation}

\noindent with

\begin{equation}
t = {c \alpha N_c\over \pi Q_s} e^{\left({\sqrt{{2\pi\over c \alpha N_c}}}-\zeta\right)}= t_1 e^{-\zeta}.
\end{equation}

\noindent So long as ${{d\over d\zeta}(\ell_\perp/Q_s)\over (\ell_\perp/Q_s)}$ is small the approximation of having
only gluons of equal rapidity interact is good.  This is perhaps most easily seen by imagining that at time \ $t$\
a particular gluon begins to cross, making a finite angle with the z-axis, the gluon medium of thickness $\Delta z
\approx 2t.$  If our particular gluon is right-moving, then during a time \ $t$\ it will cross all the left-moving
gluons in the medium.  If during that time, corresponding to $\Delta \zeta \approx 1,$ the gluon does not suffer
enough interactions to change its momentum by a significant amount, then its important interactions must occur
much later and with gluons having a similar rapidity since only such gluons have the same longitudinal velocity
and so do not separate from our particular gluon which is essentially moving along a straight line trajectory.
Now, using (64) and (65), one finds

\begin{equation}
{{d\over d\zeta}(\ell_\perp/Q_s)\over (\ell_\perp/Q_s)} = {{1\over 2}\ {d\over d\zeta}(\ell_\perp/Q_s)^2\over
(\ell_\perp/Q_s)^2}\ =\ {1\over 2\zeta}.
\end{equation}

\noindent Thus so long as $\zeta$ is large our procedure of calculation should be reasonable.  However, somewhere
when $\zeta$ becomes on the order of 1 our whole procedure begins to break down as scattering between unequal
rapdity gluons becomes important.  In addition, at this time, scatterings which transform longitudinal momentum into
transverse momentum are also becoming important, and they are not included in our formalism.

Thus when $\zeta$ is of order 1 we expect the interactions important for equilibration to begin and our
approximation to break down.  As can be seen by comparing (58) and (63) this is essentially the same criterion we
found in the last section.  Thus, in our simple approach we are not able to follow gluons when the interactions
essential for equilibration begin.  We are, however, able to follow the system, on average, up to the time at
which equilibrating interactions begin to occur.  We have a good estimate of that time which is close to
$t_1,$ given in (53), and of the transverse momentum of the gluons, given by (59), at the time when equilibration
starts.  However, because the density of gluons, given by (18), is of size $c{N_c^2-1\over 4\pi^2\alpha N_c} Q_s^3 e^{-\xi_1}$ 
when equilibration starts we expect that when equilibration is complete and ${dN\over dx} \sim \ell_\perp^3$ 
that $(\ell_\perp/Q_s)^2 \sim e^{-{2\over 3}\ \xi_1}.$  It is encouraging that equilibration starts
well before the expansion changes from one-dimensional to three-dimensional.

\bigskip
\noindent{\bf Acknowledgement}

I am grateful to Professor Larry McLerran for emphasing to me the importance of the small angle gluon-gluon
scatterings in this problem.  This observation was crucial for the discussion given above.  I would like to thank Dr. Dominique Schiff and the Laboratoire
de Physique Theorique, where this work was completed, for their hospitality.

\end{document}